\documentstyle[twocolumn,psfig,floats,aps,prl,amsmath,amssymb]{revtex}
\begin{document}
\twocolumn[\hsize\textwidth\columnwidth\hsize\csname
@twocolumnfalse\endcsname
\title{Study of impurities in spin-Peierls systems including 
lattice relaxation}
\author{P.~Hansen$^a$, D.~Augier$^b$, J.~Riera$^{a,b}$ and D.~Poilblanc$^b$
}
\address{
$^a$Instituto de F\'{\i}sica Rosario, Consejo Nacional de 
Investigaciones 
Cient\'{\i}ficas y T\'ecnicas y Departamento de F\'{\i}sica\\
Universidad Nacional de Rosario, Avenida Pellegrini 250, 2000-Rosario,
Argentina\\
$^b$Laboratoire de Physique Quantique \& Unit\'e mixte
de Recherche CNRS 5626\\
Universit\'e Paul Sabatier, 31062 Toulouse, France
}
\date{\today}
\maketitle
\begin{abstract}

The effects of magnetic and non-magnetic impurities 
in spin-Peierls systems are investigated 
allowing for lattice relaxation and quantum fluctuations.
We show that, in isolated chains, strong bonds form next to impurities,
leading to the appearance of magneto-elastic solitons. 
Generically, these solitonic excitations do not bind to impurities. 
However, interchain elastic
coupling produces an attractive potential at the
impurity site which can lead to the formation of bound states.
In addition, we predict that small enough chain segments 
do not carry magnetic moments at the ends.


\end{abstract}

\vskip2pc]

Quasi-one dimensional (1D) spin-Peierls systems attract intense experimental
and theoretical activity for their fascinating magnetic properties.
Such systems usually consist in weakly coupled spin-1/2 Heisenberg chains. 
Due to spin-phonon coupling, 
these materials undergo at low temperature a transition towards a phase
exhibiting a lattice dimerization and a spin gap~\cite{SP}.
Inorganic compounds like CuGeO$_3$ are easily doped
by magnetic or non-magnetic impurities by substituting 
a fraction of the spin-1/2 Cu$^{2+}$ ions by spin-0 Zn$^{2+}$ or
spin-1 Ni$^{2+}$ ions. 
As shown by magnetic susceptibility measurements~\cite{suscep}
and inelastic neutron scattering experiments~\cite{neutrons},
doping with impurities leads to a rapid collapse of the spin gap. 
Competition between the spin-Peierls phase and a new antiferromagnetic 
(AF) phase induced by doping has been established by magnetic
susceptibility measurements~\cite{suscep,suscep2},
specific heat measurements~\cite{specific_heat}, neutron 
scattering~\cite{Regnault} and NMR experiments~\cite{Renard}. 
These experiments suggest that magnetic moments and 
enhanced staggered spin correlations are induced by impurity doping.

 From a theoretical point of view, the relevant phonons in 
dimerized quasi-1D compounds like 
CuGeO$_3$ or NaV$_2$O$_5$ are often considered as three-dimensional,
an assumption which, {\it a priori}, would justify a classical
treatment of the lattice.
In the dimerized AF Heisenberg chain, a model
widely used in the literature to describe these materials, 
one introduces a fixed dimerization $J(1\pm\delta)$ of the
magnetic exchange integral leading to the opening of a spin gap 
$\propto \delta^{2/3}$. In CuGeO$_3$, it is believed that magnetic
frustration (i.e. AF coupling between next nearest
neighbor sites) plays a role.\cite{rieradobry,castilla}
The lowest energy excitations of the dimerized Heisenberg chain
consist of spinon-spinon bound states~\cite{bound_states,Sorensen} 
lying below the two-magnon continuum. 

Extensive work on the effect of impurities in dimerized spin
chains have been carried out~\cite{Martins,Laukamp,Fukuyama}.
The introduction of vacancies creates finite chains.
So far, those studies ignore the
lattice dynamics which, physically, is justified only
when the elastic coupling to the neighboring chains
is large enough. In this case, a finite chain can end by
either a ``weak'' or a ``strong'' bond depending on the sign of 
the dimerization on this bond. These two types of boundaries show
very different magnetic properties: in contrast to the strong bond 
edge, a weak bond edge can localize a $S=1/2$ magnetic excitation.
This effect is responsible for the presence of strong AF
correlations in the vicinity of weak edges.\cite{Fabrizio}

In the approach discussed above, the effects of impurities have been 
considered under the assumption of a space- and time-independent lattice 
dimerization. 
However, due to the magneto-elastic coupling, the presence of a 
spin-1/2 excitation is expected to, locally, distort the 
underlying lattice creating an elastic soliton~\cite{soliton}.
Such effects were recently investigated 
in the context of the incommensurate phase of spin-Peierls systems under 
magnetic field~\cite{IC,Kiryukhin}.
In this Letter, 
we investigate lattice relaxation effects in Heisenberg chains 
in the vicinity of spin-0 or spin-1 impurities 
by exact diagonalization (ED) and quantum Monte Carlo (QMC) simulations.
The lattice is treated either classically in the adiabatic 
approximation~\cite{IC}, i.e. allowing for non-uniform static lattice 
distortions, or in a fully quantum mechanical way~\cite{Augier,Wellein} 
introducing, in addition, the lattice dynamics. 
In an isolated chain, we have found that strong bonds form
next to an impurity. For nonmagnetic impurities, solitonic
excitations do not bind to the impurity.
However, in this case, the interchain elastic
coupling generates an effective impurity--soliton attractive potential
which leads to the formation of a bound state.
The spatial extension of this bound state is governed by the 
strength of the interchain coupling.  

The model we first consider is purely 1D and includes a classical
lattice distortion, 
\begin{eqnarray}
   {\cal H_\parallel}=
   J\sum_{i} (1+\delta_{i})
   {\bf S}_{i} \cdot {\bf S}_{i+1} 
+ \frac{1}{2}K \sum_{i} \delta_{i}^2 \, ,
\label{ham} 
\end{eqnarray}
\noindent
where the second part corresponds to the elastic energy 
lost within the chain. The role of the interchain elastic
coupling ${\cal H}_\perp$ will be discussed later. 
The classical bond modulations $\delta_i$ 
have to be determined from a minimization of the total energy.
The pure system is dimerized with $\delta_i=\delta (-1)^i$.

\begin{figure}[htb]
\begin{center}
\vspace{-0.2truecm}
\psfig{figure=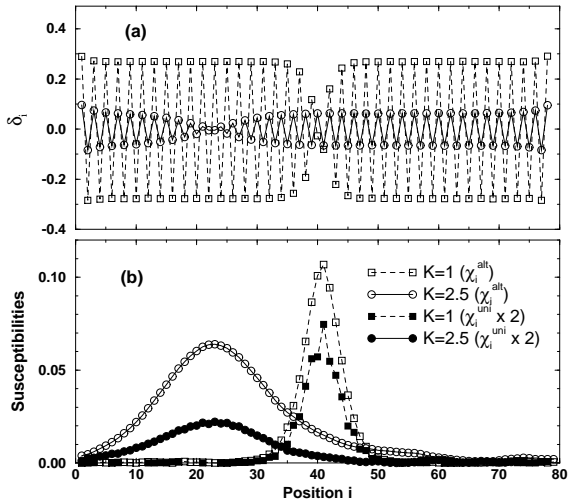,width=8truecm,angle=-90}
\end{center}
\caption{
QMC calculation (at $T=0.05$) of the 
modulation $\delta_i$ (a) and the local susceptibility
$\chi_i$ (b) (decomposed into its uniform $\chi^u_i$ and alternating 
$\chi^a_i$ components) on a $L=79$ open chain. Data for $K=1$ and
$K=2.5$ are shown.
}
\label{Pattern1}
\end{figure}

We investigate the effect of spin-0 impurities 
by considering open chains.
Such chains with an odd number of sites must contain at least a
spin-1/2 excitation. 
Previous QMC simulations supplemented by a
self-consistent determination of the equilibrium 
distortion pattern $\delta_i$ (see Ref.~\cite{IC} for details) 
have been extended to this new physical situation.
Results for a $L=79$ sites open chain shown in Fig.~\ref{Pattern1}
for various parameters reveal the existence of a single
solitonic excitation located away from the chain edges. 
The zero temperature local susceptibility 
$\chi_i=\sum_j\big< S_{i}^ZS_j^Z\big>$ at site 
$i$ corresponds physically to the average value of $S_i^Z$ {\it with
respect} to the global orientation of the total $S^Z$ spin 
component. \cite{eggert}
Fig.~\ref{Pattern1}(b) show that $\chi_i$ oscillates rapidly between 
positive and negative values (large staggered component) 
and has the largest amplitude
of both its uniform and staggered components in the region 
where the dimer order parameter is suppressed, i.e. around the
soliton. This result is quite different to that seen in fixed
dimerization calculations\cite{Laukamp} where it was observed
that spin-1/2 excitations are bound to the chain edge.
For increasing $K$ (i.e. for decreasing spin-lattice coupling),
the width of the soliton increases and the solitonic pattern 
continuously evolves into a sinusoidal distortion as expected in the weak 
coupling limit. It should be stressed that different QMC runs lead to random
degenerate equilibrium 
solitonic patterns centered on different sites in a wide area around 
chain  center.
However, no changes occur at the two edges of 
the chains which, systematically,
end with a strong bond (i.e. $\delta_i>0$).
This clearly indicates that there exists a short range repulsion 
between the impurity (edges) and the soliton. 
One should notice that, independently of the presence of the soliton
(i.e. for odd or even chains), $|\delta_i|$ increases in the close 
vicinity of the edges, in contrast with the analytical results of 
Ref.~\cite{Fukuyama}.

\begin{figure}[htb]
\begin{center}
\vspace{-0.2truecm}
\psfig{figure=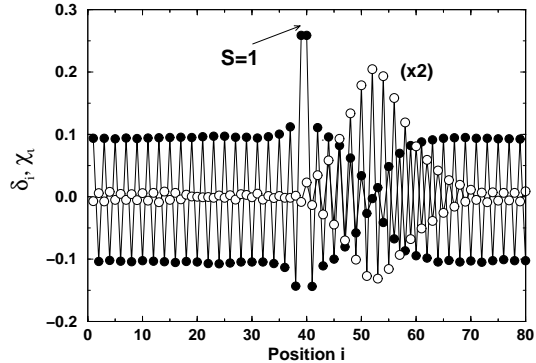,width=7truecm,angle=0}
\end{center}
\caption{
Monte Carlo calculation (at $T=0.05$) of the 
modulation $\delta_i$ ($\bullet$) and the local susceptibility $\chi_i$ 
($\circ$) on a $L=80$ closed chain, $K=2$, 
and a spin-1 impurity at site $i=40$.
}
\label{Pattern2}
\end{figure}

The case of a spin-1 impurity has also been 
considered by assuming, for simplicity, 
the same exchange integral and elastic 
constant on the two bonds on each side of the impurity.
Previous calculations~\cite{Hansen} assuming a uniform dimerization of the
chain have shown that spin-1 impurities lead to more localized 
states than spin-0 impurities (static vacancies). 
The QMC results shown in Fig.~\ref{Pattern2} reveal that the 
two bonds next to the spin-1 impurity become especially 
strong indicating that the impurity and the two
neighboring S=1/2 spins form an effective spin-0 defect 
leading qualitatively to the same physics as in the
case of a $S=0$ vacancy.
Indeed, the solitonic pattern shown in Fig.~\ref{Pattern2}(b) resembles the
ones obtained previously. However, it should be noticed that the
new profile is not completely symmetric and that the soliton is always 
located close to the impurity. This signals a small attraction
in the vicinity of a spin-1 impurity. This might be due to the fact
that the three-site system formed by the $\rm S=1$ impurity and its
two spin $\rm S=1/2$ neighbors spends most of the time ($\approx
80 \%$) but {\it not all the time} in the $S=0$ state.

In order to investigate the role of the lattice dynamics, next we 
generalize the previous approach by assuming a coupling
to dynamical phonons,
\begin{eqnarray}
   {\cal H_\parallel}=
   J\sum_{i} (1+g(b_i+b_i^\dagger))
   {\bf S}_{i} \cdot {\bf S}_{i+1} 
+\Omega\, b_i^\dagger b_i  \,  .
\label{ham_dyn} 
\end{eqnarray}
\noindent
For sake of simplicity, optical dispersionless modes are considered here. 
While the adiabatic treatment discussed above is justified in the
$\Omega\rightarrow 0$ limit, the lattice dynamics can not be
neglected when $\Omega$ and $J$ becomes comparable. 
In this case, the lattice modulation can be defined by
$\delta_i=g\big< b_i+b_i^\dagger\big>$.
The treatment of the phononic degrees of freedom will
relie here on a variational approach~\cite{Fehrenbacher} which gives
accurate results~\cite{Augier}. 
Previous calculations~\cite{Augier} have shown that the lowest 
$S=1/2$ excitations of this model correspond also to 
massive solitons and antisolitons. Furthermore, soliton and 
antisoliton do not bind in the strictly 1D case.

\begin{figure}[htb]
\begin{center}
\vspace{-0.4truecm}
\psfig{figure=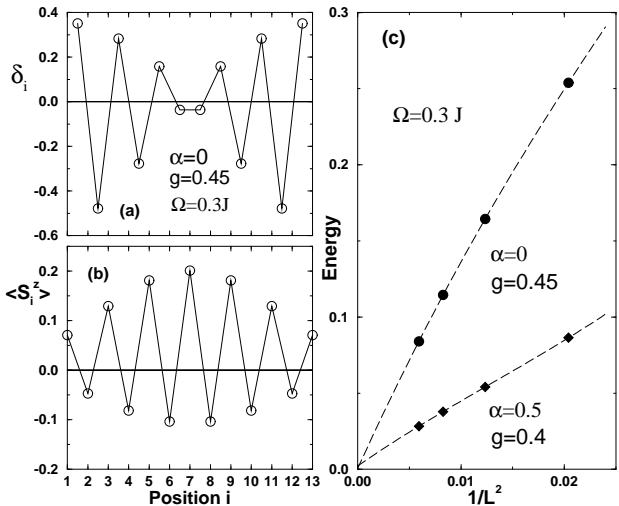,width=8truecm,angle=-90}
\end{center}
\caption{
Modulation $\delta_i$ (a) and spin density $\big< S^Z_i\big>$ (b) in the GS of
a $L=13$ sites open Heisenberg chain coupled to dynamical phonons. 
Parameters are shown on the plot. 
(c) Finite size scaling of the soliton-impurity binding energy
in the case of a vacancy (as indicated on the plot).  
}
\label{Pattern3}
\end{figure}

The results obtained by ED for an open chain 
with an odd number of sites are shown in
Fig.~\ref{Pattern3}. The lattice modulation pattern and the 
spatial variation of the spin polarization are very similar to
the results obtained in the adiabatic treatment of the lattice. 
In particular, strong bonds ($\delta_i>0$) also form at the chain 
ends and a soliton appears in the middle of the chain.

In order to get information on the effective interaction
between a soliton and the chain ends, it is instructive
to define the soliton-impurity binding energy
on a $L=2p+1$ chain as
$ E_B (L) = E_{IS}(L)-E^*_{0}(L) - e_S - e_I$,
where $E_{IS}(L)$ is the ground state energy of an $L$-site chain
with an impurity,
$E^*_{0}(L)$ is the energy of the pure system obtained for even
number of sites and interpolated to $L$,
and $e_S~(e_I)$ is the extrapolated soliton (impurity) energy 
(see Ref.~\cite{Augier}).
The finite size scaling of $E_B$ is shown in Fig.~\ref{Pattern3}
and reveals no binding in the thermodynamic limit. 
The conclusion is also similar when a finite magnetic frustration 
$\alpha=J_2/J\ne 0$ is considered (in that case, a term
${\cal H}_F=J_2\sum_i {\bf S}_i\cdot{\bf S}_{i+2}$ is added to 
Hamiltonian~(\ref{ham_dyn})).
In contrast, analogous calculations for $\rm S=1$ impurities
indicate a nonzero binding between the impurity and the 
soliton.\cite{augierfut}
These results including quantum lattice fluctuations
are consistent
with the previous ones in the adiabatic approximation.
\begin{figure}[htb]
\begin{center}
\psfig{figure=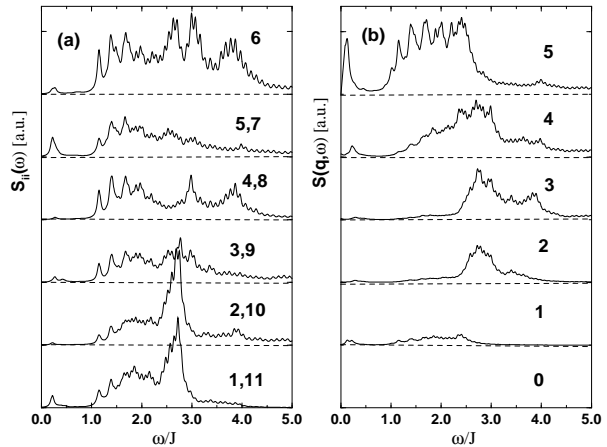,width=8truecm,angle=-90}
\end{center}
\vspace{-0.2cm}
\caption{
(a) Local dynamical spin structure factor $S_{ii}(\omega)$ 
calculated on the various sites (indicated on the plot) 
of a $L=11$ spin chain coupled to dynamical phonons. 
(b) Dynamical spin structure factor $S(q,\omega)$ for the same system
as (a) (q in units of $2\pi/L$). Parameters are as in 
Fig. \ref{Pattern3}.
}
\label{Dynamics}
\end{figure}

The local dynamical spin-spin correlation function is shown
in Fig.~\ref{Dynamics}(a). The reminiscence of the spin gap
of the pure chain is clearly seen at an energy $\omega\sim J$.
However, spectral weight appears at much lower energy.
It can be attributed to the soliton excitation which behaves as a
$S=1/2$ object weakly connected to the rest of the system,
consistently with the behavior discussed above (Fig. \ref{Pattern1}).
If one labels the sites from 1 to $L$ starting from
the left end of the chain, we observe a large low energy weight
at the ``odd'' positions, $i=2k+1$ due to the dimerization pattern,
with the largest peak at the closest site to the center. In fact, 
this can be qualitatively 
understood by assuming that the (free) soliton can move 
by hopping ``over'' a strong bond. 
This feature also manifests itself in the low energy peak in
$S({\bf q},\omega)$ near ${\bf q}=\pi$ as it can be seen in 
Fig. ~\ref{Dynamics}(b). This is similar to what has been observed in
fixed dimerization calculations.\cite{Martins}
The remnants of high energy branch
of the pure system can be still seen.

\begin{figure}[htb]
\begin{center}
\psfig{figure=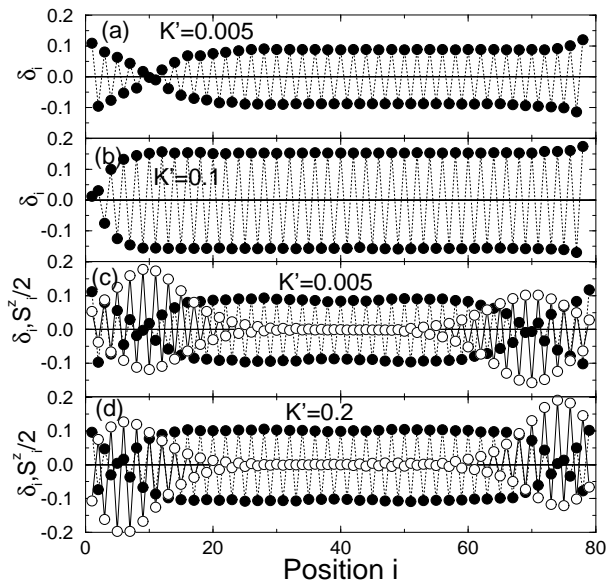,width=9truecm,angle=00}
\end{center}
\vspace{-0.2truecm}
\caption{
(a), (b): QMC calculation (at $T=0.05$) of the 
modulation $\delta_i$ vs $i$ on a $L=79$ chain with OBC
coupled to a small external dimerization.
(c), (d): the same for a $L=80$ chain. In this case $S^z_i$
vs $i$ is also shown.
}
\label{Pattern4}
\end{figure}

Lastly, we investigate the role of a realistic interchain coupling 
${\cal H}_\perp$. 
To illustrate the role of ${\cal H}_\perp$, let us consider 
the physical situation of a finite chain cut by two spin-0 impurities
at its ends and immersed in the bulk.
In the dimerized phase, the neighboring 
chains produce a $q=\pi$ potential of the form,
$ {\cal H}_\perp=K^\perp \sum_i \delta_i \delta^{ext}_i$.
Here the modulation of the neighboring chains is treated in
the mean-field approximation,
i.e. $\delta^{ext}_i = (-1)^i \delta_{0}$,
but the full spatial dependence
in the chain with the impurities is retained.
The amplitude of the external potential is then related to
the elastic constant $K_\perp$ between the chains by 
$K^\prime=K_\perp\delta_{0}$.
In the case of finite chains with an odd number of sites,
the external potential tends to form a weak bond on, let say, the left
end. Therefore, the soliton will experience a confining force 
proportional to its separation from the left end.
This attractive potential originates physically from the misfit 
between the dimerization pattern on the left side of the soliton and
the dimerization pattern of the bulk. 
Our numerical calculations shown in Fig.~\ref{Pattern4} 
confirm this intuitive picture. 
The equilibrium position is obtained when the
small confining potential is equilibrated by the short range repulsive
potential created by the impurity. 

The case of an open chain with an even number of sites is also 
particularly interesting (Fig.~\ref{Pattern4}(c),(d)). 
If the external potential is 
out-of-phase with the open chain dimerization, the
external potential will lead to the formation of a 
soliton-antisoliton ($s\bar{s}$) pair
in the center of the chain.  For increasing $K^\prime$, the two 
$S=\pm 1/2$ excitations migrate towards the chain ends forming 
two localized excitations. 
These calculations again support the fact that 
soliton-impurity bound states are stabilized by the interchain 
elastic coupling.\cite{Khomskii}
Because of the finiteness of the energy cost  
associated with the formation of the $s\bar{s}$ pair, one can,
on general grounds, deduce the existence of a critical value
$K'_c$ of $K'$ in such a phenomenon. A comparison between the
energy cost $\propto 2e_S$ and the 
transversal elastic energy gain $\propto K'(L- 2 \Gamma)$, 
where $\Gamma$ is half the soliton width, leads to 
$K'_c \sim 2e_S/(L- 2 \Gamma)$. 
Our numerical calculation confirms this prediction.
Alternatively, this implies that, when two impurities are sufficiently
close to each other, the soliton and the anti-soliton bound to each of
them can annihilate each other leading to the disappearance of the
magnetic moments.
This situation would be more likely to appear for larger impurity 
concentration, i.e. for short chain length on average. 
Note that such features are almost
entirely missed in fixed dimerization calculations.

IDRIS (Orsay, France) is acknowledged for allocation of CPU-time
on the CRAY supercomputers. 

{\it Note added.}-- After completion of this paper
we learnt of a related calculation on spinon
confinement\cite{uhrig}.


\begin{references}
\vspace{-1.2cm}

\bibitem{SP} M.~Hase, I.~Terasaki and K.~Uchinokura, 
Phys. Rev. Lett.~{\bf 70} 3651 (1993).

\bibitem{suscep} M.~Hase {\it et al.},
 Phys. Rev. Lett. {\bf 71}, 4059 (1993). 

\bibitem{neutrons} J.~G.~Lussier, S.~M.~Coad, D.~F.~McMorrow 
and D.~McK.~Paul, J. Phys. Cond. Matt. {\bf 7}, L325 (1995). 

\bibitem{suscep2} T.~Masuda {\it et al.},
Phys. Rev. Lett. {\bf 80}, 4566 (1998);
K.~Manabe {\it et al.}, cond-mat/9805072.

\bibitem{specific_heat} S.~B.~Oseroff {\it et al.},
 Phys. Rev. Lett. {\bf 74}, 1450 (1995).

\bibitem{Regnault} L.-P.~Regnault {\it et al.}, Phys. Rev. B
  {\bf 53}, 5579 (1996).

\bibitem{Renard} J.-P.~Renard {\it et al.}, Europhys. Lett. 
 {\bf 32}, 579 (1995).

\bibitem{rieradobry} J. Riera and A. Dobry, Phys. Rev. B {\bf 51},
            16098 (1995).

\bibitem{castilla} G. Castilla, S. Chakravarty and V.J. Emery, Phys.
      Rev. Lett. {\bf 75}, 1823 (1995).

\bibitem{bound_states} G.~S.~Uhrig and H.~J.~Schulz,
Phys. Rev. B {\bf 54}, R9624 (1996); A.~Fledderjohann and C. Gros,
Europhys. Lett. {\bf 37}, 189 (1997);
D.~Augier, D.~Poilblanc, S.~Haas, A.~Delia, E.~Dagotto,
Phys. Rev. B {\bf 56}, R5732 (1997). 

\bibitem{Sorensen} E.~Sorensen {\it et al.}, cond-mat/9805386,
Phys. Rev. B, in press.

\bibitem{Martins} G.~B.~Martins, E.~Dagotto and J.~Riera,
Phys. Rev. B {\bf 54}, 16032 (1996).

\bibitem{Laukamp} M. Laukamp {\it et al.}, Phys. Rev. B {\bf 57},
10755 (1998). 

\bibitem{Fukuyama} H. Fukuyama, T. Tanimoto, and M. Saito, J. Phys. 
      Soc. Jpn. {\bf 65}, 1182 (1996).

\bibitem{Fabrizio} M. Fabrizio and R. M\'elin, Phys. Rev. B {\bf 56},
    5996 (1997).

\bibitem{soliton} S.~A. Brazovskii, S.~A. Gordynin and 
N.~N. Kirova, JETP Lett. {\bf 31}, 456 (1980);
M. Fujita and K.~Machida, J. Phys. Soc. Jpn. {\bf 53}, 4395 (1984).

\bibitem{IC} A. Dobry and J. Riera, Phys. Rev. B {\bf 56}, 2912 (1998);
 A.~E.~Feiguin {\it et al.}, Phys. Rev. B {\bf 56}, 14607
(1997); F. Sch\"onfeld {\it et al.}, cond-mat/9803084. 

\bibitem{Kiryukhin} For related experiments see e.g., V. Kiryukhin
 {\it et al.}, Phys. Rev. Lett. {\bf 76}, 4608 (1996).

\bibitem{Augier} D.~Augier {\it et al.}, Phys. Rev. B {\bf 58}, 9110 (1998).

\bibitem{Wellein} G.~Wellein, H.~Feske and A.~P.~Kampf,
cond-mat/9804085, Phys. Rev. Lett., in press.

\bibitem{eggert} S. Eggert and I. Affleck, Phys. Rev. Lett. {\bf 75},
      934 (1995).

\bibitem{Hansen} P. M. Hansen {\it et al.}, Phys. Rev. B {\bf 58},
6258 (1998). 

\bibitem{Fehrenbacher} R. Fehrenbacher, Phys. Rev. Lett. {\bf 77},
     2288 (1996).

\bibitem{augierfut} D. Augier {\it et al.}, cond-mat/9807265.

\bibitem{Khomskii}D. Khomskii, W. Geertsma, and M. Mostovoy, Czech.
 J. Phys., {\bf 46}, Suppl. S6, 3239  (1996).

\bibitem{uhrig} G.~S.~Uhrig et al., cond-mat/9805245. 
In contrast to the present work, these authors 
do not include lattice relaxation around impurities.

\end{references}
\end{document}